\documentstyle[aps,prl,twocolumn]{revtex}

\begin{document}

\noindent
{\bf Comment on ``Density of States of Disordered Two-Dimensional Crystals
with Half-Filled Band''}
  
In a recent letter \cite{nko} Nakhmedov et al. (NKO) claimed that the
Van Hove
singularity at $\epsilon=0$ in the density of states (DoS) $\rho(\epsilon)$
of the two-dimensional
crystal with half-filled tight-binding band
survives the addition of substitutional impurities.  The authors
found a singular average DoS as ($\epsilon \to 0$)
(Eq. (14)
in Ref. \cite{nko}):
\begin{equation}
\rho(\epsilon) \propto (2\pi \epsilon_F \tau_\pi)^{1/2} 
\ln^{1/2} \left( \frac{1}{4\tau_\pi |\epsilon|} \right), 
\label{DoS}
\end{equation}
where $\epsilon_F$ is the Fermi energy and $\tau_\pi$ is
the electronic relaxation time due to elastic processes where
incoming and outgoing momenta differ by a nesting wave vector
${\bf P}_0=(\pi,\pi)$.
This result is very counter-intuitive. Naively, one would expect that the
potential disorder simply broadens local energy levels and washes out  any
singularities in the DoS. 
In this comment we show that the DoS is indeed
non-singular. The derivation in Ref. \cite{nko} suffers from several
inconsistencies. We proceed by correcting them one by one, and find a finite DoS
at half filling. 

First, the authors used the standard impurity averaged perturbation theory,
and calculated the weak localization correction to the DoS. This correction was
added to the DoS $\rho_0^{(2)}(\epsilon)$ of the pure crystal. This is
inconsistent with the standard scheme that the authors use \cite{agd} in which
one starts with the self-consistent Born approximation for Green's functions 
and the DoS:
\begin{equation}
\rho_B (\epsilon) = - \frac{2}{\pi} \, \text{Im} \! \int \! 
\frac{d^2 p}{(2\pi)^2} \, G_R^0({\bf p}, \epsilon)\, , 
\end{equation}
where $G_R^0({\bf p}, \epsilon) = \{\epsilon - [\epsilon({\bf p}) -
\epsilon_F] + \frac{i}{2\tau}\}^{-1}$ is the retarded Green's function.
Evaluating this integral for energies smaller than $1/\tau$ we obtain a
non-singular DoS, since in this case the Van Hove singularity is cut off by the
scattering rate:   
\begin{equation}
\rho_B (0) = \frac{2}{(\pi a)^2 \epsilon_F} \, \ln (\epsilon_F \tau)\, .
\label{DoSBorn}
\end{equation}

Next, we consider the $1/(\epsilon_F \tau)$ correction to this DoS. In
Ref. \cite{nko} it was found to be singular. In their analysis the authors
separated the scattering processes into two classes --- with momentum transfers
close to 0 and the nesting vector ${\bf P}_0$. The scattering with nesting was
assumed to be dominant, and the corresponding scattering times, $\tau_0$ and
$\tau_\pi$, very different: $\tau_\pi \ll \tau_0$. This statement does not
agree with the expressions for these times given in Ref. \cite{nko}:
\begin{eqnarray*}
\frac{1}{\tau_0} &=& \frac{C_{\rm imp}}{(2\pi)^2} \int 
\frac{d{\bf S}}{|{\bf V}_{\bf k}|} |U({\bf k},0)|^2\, , \\
\frac{1}{\tau_\pi} &=& \frac{C_{\rm imp}}{(2\pi)^2} \int 
\frac{d{\bf S}}{|{\bf V}_{\bf k}|} |U({\bf k},{\bf P}_0)|^2\, ,
\end{eqnarray*}
where $C_{\rm imp}$ is the concentration of impurities,  
${d{\bf S}}/{|{\bf V}_{\bf k}|}$ is the usual Fermi surface measure, and
$U({\bf k},{\bf q})$ is the scattering matrix element for momentum transfer
${\bf q}$.   For example, in the case of point-like impurities ($U({\bf
k},{\bf q}) = {\rm const}$) one finds $\tau_\pi = \tau_0$, while for reasonable
finite size impurities these times are of the same order of magnitude. However,
from purely formal point of view one still can consider the model with
$\tau_\pi \ll \tau_0$.

Finally, the weak localization correction to the DoS is expressed in
terms of the quantity $\alpha(\epsilon)$, see Eqs. (9,10) in Ref. \cite{nko},
which involves integrals of Cooperons and diffusons, Eqs. (5,6) of Ref. \cite{nko}.
Evaluating these integrals, NKO cut off the leading logarithmic contribution at $k l \sim
\sqrt{\epsilon \tau}$. 
This is correct for energies in the range  
$\epsilon \tau_0 > 1$. However, when $\epsilon \to 0$, the logarithm is cut off
by any finite $\tau_0$, since then $1 - (\tau/\tau_ \pi)^2$ in the denominators
of NKO's Eqs. (5,6) is small but finite.  Therefore, the quantity
$\alpha_0(\epsilon)$ saturates as $\epsilon \to 0$ at the finite value
\begin{equation}
\alpha_0(0) = \frac{1}{2\pi \epsilon_F \tau_\pi} 
\ln \frac{\tau_0}{\tau_\pi}\, . 
\label{alpha}
\end{equation}
Since the expansion parameter $1/\epsilon_F \tau_\pi \ll 1$,
the weak localization correction is also small, $\alpha_0(0) \ll 1$, and does
not essentially modify the DoS $\rho_B (0)$.
The value $\alpha_0(0) \sim 1$ is only achieved for $\tau_0/\tau_\pi \sim
\exp(2\pi \epsilon_F \tau_\pi)$. Such extremely large value cannot be obtained
for a generic disorder potential. 

Combining our Eqs. (\ref{DoSBorn},\ref{alpha}) 
with Eq. (12) in Ref. \cite{nko}, we arrive at 
\begin{equation}
\rho(0) = \frac{2}{(\pi a)^2 \epsilon_F} 
\frac{\ln (\epsilon_F \tau)}{\sqrt{1 + \alpha_0(0)}}\, ,
\end{equation}
a finite result. 

{\it Acknowledgement.} PJH thanks A. Yashenkin for useful discussions, and
the Institute for Theoretical Physics for support through  NSF PHY94-07194.
This research was supported in part by 
%the National Science Foundation under Grant No. 
NSF PHY99-07949.    
                                     
\vskip5mm

\noindent
Ilya A. Gruzberg$^1$, P. J. Hirschfeld$^2$, and A. V. Shytov$^1$.

\vskip3mm

$^1$Institute for Theoretical Physics

$\phantom{^1}$University of California

$\phantom{^1}$Santa Barbara, CA 93106-4030

$^2$Physics Department

$\phantom{^2}$University of Florida

$\phantom{^2}$Gainesville, FL 32611

\end{document}